\documentclass[runningheads]{svmult}

\usepackage{makeidx}   % allows index generation
\usepackage{graphicx}  % standard LaTeX graphics tool
\usepackage{subeqnar}  % subnumbers individual equations
\usepackage{multicol}  % used for the two-column index
\usepackage{physprbb}  % modified textarea for proceedings,
\makeindex             % used for the subject index
\newcommand{\be}{\begin{equation}}\newcommand{\ee}{\end{equation}}
\newcommand{\bea}{\begin{eqnarray}}\newcommand{\eea}{\end{eqnarray}}
\newcommand{\brr}{\begin{array}}\newcommand{\err}{\end{array}}
\newcommand{\bit}{\begin{itemize}}\newcommand{\eit}{\end{itemize}}
\newcommand{\ben}{\begin{enumerate}}\newcommand{\een}{\end{enumerate}}

\newcommand{\ba}{\begin{array}}
\newcommand{\ea}{\end{array}}

\def\noi{\noindent}

\def\1{{_{1}}}
\def\2{{_{2}}}
\newcommand{\IC}{{\textsc C}\hspace{-2mm}{\textsl I}}
\newcommand{\IR}{{\textsc I}\hspace{-0.6mm}{\textsc R}}

\begin{document}
\title*{Information theory and generalized statistics}
\toctitle{Information theory and generalized statistics}
% \protect\newline }
% allows explicit linebreak for the table of content
%
%
\titlerunning{Information theory and generalized statistics}
% allows abbreviation of title, if the full title is too long
% to fit in the running head
%
\author{Petr Jizba}
\authorrunning{P.~Jizba}
% if there are more than two authors,
% please abbreviate author list for running head
%
%
\institute{Institute of Theoretical Physics, University of
Tsukuba, Ibaraki 305--8571, Japan }

\maketitle              % typesets the title of the contribution
\vspace{-4mm}
\begin{abstract}
In this lecture
%\index{abstract}
we present a discussion of generalized statistics based on
R\'{e}nyi's, Fisher's and Tsallis's measures of information. The
unifying conceptual framework which we employ here is provided by
information theory. Important applications of generalized
statistics to systems with (multi--)fractal structure
% long--range
%interactions and long--time memories
are examined.
\end{abstract}

%%%%%%%%%%%%%%%%%%%%%%%%%%%%%%%%%%%%%%%%%%%%%%%%%%%%%%%%%%%%%%%%%%%%
\section{Introduction}
%%%%%%%%%%%%%%%%%%%%%%%%%%%%%%%%%%%%%%%%%%%%%%%%%%%%%%%%%%%%%%%%%%%%
%
One of the important approaches to statistical physics is provided
by information theory erected by Claude Shannon in the late 1940s.
Central tenet of this approach lies in a construction of a measure
of the ``amount of uncertainty" inherent in a probability
distribution~\cite{Sh1}. This measure of information (or Shannon's
entropy) quantitatively equals to the number of binary (yes/no)
questions which brings us from our present state of knowledge
about the system in question to the one of certainty. The higher
is the measure of information (more questions to be asked) the
higher is the ignorance about the system and thus more information
will be uncovered after an actual measurement. Usage of Shannon's
entropy is particularly pertinent to the Bayesian statistical
inference where one deals with the probability distribution
assignment subject to prior data one possess about a
system~\cite{Jay1,Jay2}. Here the prescription of maximal
Shannon's entropy (i.e., maximal ignorance) - MaxEnt, subject to
given constraints yields the least biased probability distribution
which naturally protects against conclusions which are not
warranted by the prior data. In classical MaxEnt the
maximum--entropy distributions are always of an exponential form
and hence the name generalized canonical distributions. Note that
MaxEnt prescription, in a sense, resembles the principle of
minimal action of classical physics as in both cases extremization
of a certain functionals - the entropy or action functionals -
yields physical predictions. In fact, the connection between
information and the action functional was conjectured by
E.T.~Jaynes~\cite{Jay2} and J.A.~Wheeler~\cite{Wheel1}, and most
recently this line of reasonings has been formalized e.g., by
B.R.~Frieden in his``principle of extreme physical information -
EPI"~\cite{Frie1}.

On a formal level the passage from information theory to
statistical thermodynamics is remarkably simple. In this case a
maximal--entropy probability distribution subject to constraints
on average energy, or constant average energy and number of
particles yields the usual canonical or grand--canonical
distributions of Gibbs, respectively. Applicability in physics is,
however, much wider. Aside of statistical thermodynamics, MaxEnt
has now become a powerful tool in non--equilibrium statistical
physics~\cite{Zub1} and is equally useful in such areas as
astronomy, geophysics, biology, medical diagnosis and economics.
For the latest developments in classical MaxEnt the interested
reader may consult ref.~\cite{Max} and citations therein.

As successful as Shannon's information theory has been, it is
clear by now that it is capable of dealing with only a limited
class of systems one might hope to address in statistical physics.
In fact, only recently it has become apparent that there are many
situations of practical interest requiring more ``exotic"
statistics which does not conform with the generalized canonical
prescription of the classical MaxEnt (often referred as
Boltzmann--Gibbs statistics). Percolation, polymers, protein
folding, critical phenomena or stock market returns provide
examples. On the other hand, it cannot be denied that MaxEnt
approach deals with statistical systems in a way that is
methodically appealing, physically plausible and intrinsically
nonspeculative (i.e., MaxEnt invokes no hypotheses beyond the
sample space and the evidence that is in the available data). It
might be therefore desirable to inspect the axiomatics of
Shannon's information theory to find out whether some ``plausible"
generalization is possible. If so, such an extension could provide
a new conceptual frame in which generalized measures of
information (i.e., entropies) could find their theoretical
justification. The additivity of independent mean information is
the most natural axiom to attack. On this level three modes of
reasoning can be formulated. One may either keep the additivity of
independent information but utilize more general definition of
means, or keep the the usual definition of linear means but
generalize the additivity law or combine both these approaches
together.

In the first case the crucial observation is that the most general
means compatible with Kolmogorov's axioms of probability are the
so called quasi--linear means which are implemented via
Kolmogorov--Nagumo functions~\cite{Ko1}. This approach was
pioneered by A.~R\'{e}nyi~\cite{Re1}, J.~Acz{\'e}l and
Z.~Dar\'{o}czy~\cite{Aczel1} in 60s and 70s. The corresponding
measure of information is then called R\'{e}nyi's entropy. Because
the independent information are in this generalization still
additive and because the quasi--linear means basically probe
dimensionality of the sample space one may guess that this theory
should play an important r\^{o}le in classical
information--theoretical systems with a non--standard geometry,
such as fractals, multi--fractals or systems with embedded
self--similarity. These include phase transitions and critical
phenomena, chaotic dynamical systems with strange attractors,
fully developed turbulence, hadronic physics, cosmic strings, etc.

Second case amounts to a modification of the additivity law. Out
of the infinity of possible generalizations the so called
$q$--additivity prescription has found widespread utility. The
$q$--calculus was introduced by F.~Jackson~\cite{FJ1} in 20's and
more recently developed in the framework of Quantum Groups by
V.~Drinfeld~\cite{VD2} and M.~Jimbo~\cite{MJ2}. With the help of
$q$--calculus one may formalize the entire approach in an unified
manner by defining $q$--derivative (Jackson derivative),
$q$--integration (Jackson integral), $q$--logarithms,
$q$--exponentials, etc. The corresponding measure of information
is called Tsallis or non--extensive entropy. The $q$--additivity
is in a sense minimal generalization because the non--additive
part is proportional to both respective information and is
linearly parametrized by the only one ``coupling" constant. The
non--additivity prescription might be understood as a claim that
despite given phenomena being statistically independent there
still might be non--vanishing correlation between them and hence
information might get ``entangled". One may thus expect that
Tsallis entropy should be important in systems with long--range
correlations or long--time memories. One may even guess that
quantum non--locality might become crucial playground for
non--extensive statistics.

Third generalization is still not explored in the literature. It
can be expected that it should become relevant in  e.g., critical
phenomena in a strong quantum regime. The latter can be found, for
instance, in the early universe cosmological phase transitions or
in currently much studied quantum phase transitions (frustrated
spin systems or quantum liquids being examples).

The structure of this lecture is the following: In Sections
\ref{Re} we review the basic information--theoretic setup for
R\'{e}nyi's entropy. We show its relation to (multi--)fractal
systems and illustrate how the R\'{e}nyi parameter is related to
multifractal singularity spectrum. The connection of R\'{e}nyi's
entropy with Fisher information and the metric structure of
statistical manifolds (i.e., Fisher--Rao metric) are also
discussed. In Section~\ref{Tsa} the information theoretic
rationale of Tsallis entropy are presented.
%Finally, in Conclusions and Outlooks we offer some speculations on
%further development in generalized statistics.

%%%%%%%%%%%%%%%%%%%%%%%%%%%%%%%%%%%%%%%%%%%%%%%%%%%%%%%%%%%%%%%%%%%%%%%%%
\section{R\'{e}nyi's entropy \label{Re}}
%%%%%%%%%%%%%%%%%%%%%%%%%%%%%%%%%%%%%%%%%%%%%%%%%%%%%%%%%%%%%%%%%%%%%%%%%
%
R\'{e}nyi entropies (RE) were introduced into mathematics by
A.~R\'{e}nyi~\cite{Re1} in mid 60's. The original motivation was
strictly formal. R\'{e}nyi wanted to find the most general class
of information measures which preserved the additivity of
statistically independent systems and were compatible with
Kolmogorov's probability axioms.

Let us assume that one observes the outcome of two independent
events with respective probabilities $p$ and $q$. Additivity of
information then requires that the corresponding information obeys
Cauchy's functional equation
\begin{equation}
{\mathcal{I}}(pq) = {\mathcal{I}}(p) + {\mathcal{I}}(q)\, .
\end{equation}
Therefore, aside from a multiplicative factor, the amount of
information received by learning that an event with probability
$p$ took place must be
\begin{equation}
{\mathcal{I}}(p) = -\log_2 p\, . \label{hartley}
\end{equation}
Here the normalization was chosen so that ignorant's probability
(i.e. $p = 1/2$) sets the unit of information - bit. Formula
(\ref{hartley}) is known as Hartley's information
measure~\cite{Re1}. In general, if the outcomes of some experiment
are $A_1, \ldots, A_n$ with respective probabilities $p_1, \ldots,
p_n$, and if $A_k$ outcome delivers ${\mathcal{I}}_k$ bits of
information then the mean received information reads
\begin{equation}
{\mathcal{I}} = \mbox{{\textsl{g}}}^{-1}\left(\sum_{k=1}^n p_k
\mbox{{\textsl{g}}}({\mathcal{I}}_k)\right)\, . \label{hartley1}
\end{equation}
Here $\mbox{{\textsl{g}}}$ is an arbitrary invertible function -
Kolmogorov--Nagumo function. The mean defined in
Eq.(\ref{hartley1}) is the so called quasi--linear mean and it
constitutes the most general mean compatible with Kolmogorov's
axiomatics~\cite{Ko1,Hardy}. R\'{e}nyi then proved that when the
postulate of additivity for independent events is applied to
Eq.(\ref{hartley1}) it dramatically restricts the class of
possible $\mbox{{\textsl{g}}}$'s. In fact, only two classes are
possible; $ \mbox{{\textsl{g}}}(x) = cx + d $ which implies the
Shannon information measure
\begin{equation}
{\mathcal{I}}({\mathcal{P}}) = - \sum_{k=1}^n p_k \log_2 (p_k)\, ,
\label{g1}
\end{equation}
and $\mbox{{\textsl{g}}}(x) = c \ 2^{(1-q)x} + d $ which implies
\begin{equation}
{\mathcal{I}}_q({\mathcal{P}}) = \frac{1}{(1-q)}\,
\log_2\left(\sum_{k=1}^n p^q_k \right)\, , \label{g2}
\end{equation}
with $q > 0$ ($c$ and $d$ are arbitrary constants). In both cases
${\mathcal{P}} = \{ p_1, \ldots, p_n \}$. Note that for linear
$\mbox{{\textsl{g}}}$'s the quasi--linear mean turns out to be the
ordinary linear mean and hence Shannon's information is the
averaged information in the usual sense. Information measure
defined by (\ref{g2}) is called R\'{e}nyi's information measure
(of order $q$) or R\'{e}nyi's entropy. Term ``entropy" is chosen
in a close analogy with Shannon's theory  because R\'{e}nyi's
entropy also represents the disclosed information (or removed
ignorance) after performed experiment. On a deeper level it might
be said that R\'{e}nyi's entropy measures a diversity (or
dissimilarity) within a given distribution~\cite{Rao}. In
Section~\ref{fish} we will see that in parametric statistics
Fisher information plays a similar r\^{o}le. It will be shown that
the latter measures a diversity between two statistical
populations.

To find the most fundamental (and possibly irreducible) set of
properties characterizing R\'{e}nyi's information it is desirable
to axiomatize it. Various axiomatizations can be
proposed~\cite{Re1,Jiz1}. For our purpose the most convenient set
of axioms is the following~\cite{Jiz1}:

\vspace{2mm}
\begin{enumerate}
\item For a given integer $n$ and given ${\cal{P}} = \{ p_1, p_2,
\ldots , p_n\}$ ($p_k \geq 0, \sum_k^n p_k =1$),
${\cal{I}}({\cal{P}})$ is a continuous with respect to all its
arguments.

\item For a given integer $n$, ${\cal{I}}(p_1, p_2, \ldots , p_n)$
takes its largest value for $p_k = 1/n$ ($k=1,2, \ldots, n$) with
the normalization ${\cal{I}}\left( \frac{1}{2}, \frac{1}{2}\right)
=1$.
%${\cal{I}}_{\alpha}({\cal{P}})$ takes its largest value for
%$p_k = 1/n, (k = 1,2, \ldots, n)$, i.e., the gained information is
%largest when we known least about the original system.

\item For a given $q\in {\IR}$; ${\mathcal{I}}(A\cap B) =
{\mathcal{I}}(A) +
{\mathcal{I}}(B|A)$ with\\
\\
 $\mbox{\hspace{2.5cm}}{\mathcal{I}}(B|A) = \mbox{{\textsl{g}}}^{-1}
\left(\sum_k \varrho_k(q)
\mbox{{\textsl{g}}}({\mathcal{I}}(B|A=A_k))
\right)$,\\
\\
and $\varrho_k(q) = p_k^q/\sum_k p_k^q$\, (distribution
${\mathcal{P}}$ corresponds to the experiment $A$).

%For independent events, i.e., ${\cal{R}} = {\cal{P}}\times
%{\cal{Q}}=
%\{p_iq_k \}$: \\
%${\cal{I}}_{\alpha}({\cal{P}}\times {\cal{Q}}) =
%{\cal{I}}_{\alpha}({\cal{P}}) + {\cal{I}}_{\alpha}({\cal{Q}})$.
%
%\item In general case there is a continuous invertible function
%$f(x)$ such that: ${\cal{I}}_{\alpha}\partial
%f({\cal{I}}_{\alpha})/\partial
%{\cal{I}}_{\alpha} = g({\cal{I}}_{\alpha}) f({\cal{I}}_{\alpha})$.\\
%\\
%Here ${\cal{I}}_{\alpha}({\cal{P}}\cup {\cal{Q}}) =
%{\cal{I}}_{\alpha}({\cal{I}}_{\alpha}({\cal{P}}),
%{\cal{I}}_{\alpha}(\cal{Q}))$. The scaling function $g(x)$ is
%common to both ${\cal{I}}_{\alpha}({\cal{P}}\cup {\cal{Q}}),
%{\cal{I}}_{\alpha}({\cal{P}})$ and ${\cal{I}}_{\alpha}({\cal{Q}})$
%.
\item $\mbox{{\textsl{g}}}$ is invertible and positive in $[0,
\infty)$.

\item ${\cal{I}}(p_1,p_2, \ldots , p_n, 0 ) = {\cal{I}}(p_1,p_2,
\ldots , p_n)$, i.e., adding an event of probability zero
(impossible event) we do not gain any new information.
\end{enumerate}

Note particularly the appearance of distribution $\varrho(q)$ in
axiom 3. This, so called, zooming (or escort) distribution will
prove crucial is Sections~\ref{can} and \ref{Tsa}.

Further characteristics of  expressions (\ref{g2}) were studied
extensively in~\cite{Re1,Jiz1}. We list here a few of the  key
ones.

\vspace{2mm}
%\begin{enumerate}
(a)~RE is symmetric:  ${\mathcal{I}}_q(p_1, \ldots, p_n) =
{\mathcal{I}}_q(p_{k(1)}, \ldots, p_{k(n)})$\,  ;

%(b)~R.E. is normalized: ${\mathcal{I}}_q\left(\frac{1}{2},
%\frac{1}{2}\right) = 1$\, ;

(b)~RE is nonnegative: ${\mathcal{I}}_q({\mathcal{P}})\geq 0 $\, ;

%(d)~R.E. is expansible: ${\mathcal{I}}_q(p_1, \ldots, p_n) =
%{\mathcal{I}}_q(0, p_1, \ldots, p_n) = etc. $\, ;

(c)~RE is decisive: ${\mathcal{I}}_q(0,1) = {\mathcal{I}}_q(1,0)
$\, ;

(d)~For $q\leq 1$ R\'{e}nyi's entropy is concave. For $q> 1$
R\'{e}nyi's entropy in not

~~~~~pure convex nor pure concave\, ;

(e)~RE is bounded, continuous and monotonous in $q$\, ;

(f)~RE is analytic in $q\in {\IC}_{I\cup III}$ $\Rightarrow$ for
$q=1$ it equals to Shannon's entropy,

~~~~~i.e., $\lim_{q \rightarrow 1} {\mathcal{I}}_q =
{\mathcal{I}}$\, .

%(i)~For a given $q$; ${\mathcal{I}}_q(A\cap B) =
%{\mathcal{I}}_q(A) + {\mathcal{I}}_q(B|A)$ with
%%
%\begin{eqnarray*}{\mathcal{I}}_q(B|A) = g^{-1}
%\left(\sum_k \varrho_k(q) g({\mathcal{I}}_q(B|A=A_k))
%\right)\end{eqnarray*}
%
%~~~~~and $\varrho_k(q) = (p_k)^{q}/\sum_k (p_k)^{q}$\,
%(distribution ${\mathcal{P}}$ corresponds to the
%
%~~~~~experiment $A$)\, .
%\end{enumerate}

\vspace{3mm}

Despite its formal origin R\'{e}nyi's entropy proved important in
variety of practical applications. Coding theory~\cite{Aczel1},
statistical inference~\cite{TA1}, quantum mechanics~\cite{HM1},
chaotic dynamical systems~\cite{TCH1,MHJ1,KT1,HP1} and
multifractals provide examples. The rest of Section~2 will be
dedicated to applications in multifractal systems. For this
purpose it is important to introduce  the concept of renormalized
information.

%%%%%%%%%%%%%%%%%%%%%%%%%%%%%%%%%%%%%%%%%%%%%%%%%%%%%%%%%%%%%%%%%
\subsection{Continuous probability distributions -
renormalization \label{Co1}}
%%%%%%%%%%%%%%%%%%%%%%%%%%%%%%%%%%%%%%%%%%%%%%%%%%%%%%%%%%%%%%%%%

Let us assume that the outcome space (or sample space) is a
continuous $d$--dimensional manifold. It is then heuristically
clear that as we refine the measurement the information obtained
tends to infinity. Yet, under certain circumstances a finite
information can be extracted from the continuous measurement.

To show this we pave the outcome space\footnote{For simplicity's
sake we consider that the outcome space has volume $V=1$.} with
boxes of the size $l = 1/n$. This divides the $d$--dimensional
sample space into cells labelled by an index $k$ which runs from
$1$ up to $n^d$. If ${\mathcal{F}}({\mathbf{x}})$ is a continuous
probability density function (PDF), the corresponding integrated
probability distribution ${\cal{P}}_n = \{p_{nk} \}$ is generated
via prescription
\begin{equation}
p_{nk}= \int_{\textrm{\scriptsize k-th
box}}{\mathcal{F}}({\mathbf{x}})d^d{\mathbf{x}}\, . \label{PDF1}
\end{equation}
Generic form of ${\mathcal{I}}_q({\mathcal{P}}_n)$ it then
represented as
\begin{equation}
{\cal{I}}_q({\cal{P}}_n) = \mbox{divergent in}\ n + \mbox{finite}
+ o(1)\, , \label{asympt1}
\end{equation}
where the symbol $o(1)$ means that the residual error tends to $0$
for $n \rightarrow \infty$. The $finite$ part ($\equiv
{\mathcal{I}}_q({\cal{F}}) $) is fixed by the requirement (or by
renormalization prescription) that it should fulfill the postulate
of additivity in order to be identifiable with an information
measure. Incidentally, the latter uniquely fixes the divergent
part~\cite{Jiz1} as $d\log_2 n$. So we may write
\begin{eqnarray}{\mathcal{I}}_q(p_{nk}) \approx d \log_2 n + h +
o(1)\, , \label{exp0}
\end{eqnarray}
\noindent which implies that
\begin{eqnarray}
{\mathcal{I}}_{q}({\mathcal{F}}) &\equiv& h = \lim_{n\rightarrow
\infty} \left( {\mathcal{I}}_{q}({\mathcal{P}}_{nk}) - d \log_2 n
\right) \ = \ \frac{1}{(1-q)}\, \log_2 \left(\int
{\mathcal{F}}^q({\mathbf{x}})d^d{\mathbf{x}} \right)\! .
\label{eq2}
\end{eqnarray}
\noindent The latter might be generalized to piecewise--continuous
${\mathcal{F}}({\mathbf{x}})$'s (Stiltjes integration) and to
Lebesgue measurable sets~\cite{Re1}. Needless to say that
R\'{e}nyi's entropy ${\mathcal{I}}_{q}({\mathcal{F}})$ exists iff.
the integral on the RHS of (\ref{eq2}) exists.

Note that (\ref{eq2}) can be recast into a form
\begin{eqnarray}
{{\cal{I}}}_{q}({\cal{F}}) &\equiv& \lim_{n \rightarrow \infty}
({\cal{I}}_{q}({\cal{P}}_n) - {\cal{I}}_{q}({\cal{E}}_n) )\, .
\label{negentropy}
\end{eqnarray}
with ${\cal{E}}_n = \left\{\frac{1}{n^d}, \ldots, \frac{1}{n^d}
\right\}$ being the uniform distribution. Expression
(\ref{negentropy}) represents nothing but R\'{e}nyi's
generalization of the Szilard--Brillouin negentropy.

%%%%%%%%%%%%%%%%%%%%%%%%%%%%%%%%%%%%%%%%%%%%%%%%%%%%%%%%%%%%%%%%%%
\subsection{Fractals, multifractals and generalized dimension}
%%%%%%%%%%%%%%%%%%%%%%%%%%%%%%%%%%%%%%%%%%%%%%%%%%%%%%%%%%%%%%%%%%
%
%\centerline{\cblue \bf Pedestrian's introduction
%into fractals:} \vspace{3mm}
%
Aforementioned renormalization issue naturally extends beyond
simple metric outcome spaces (like ${\IR}^d$). Our aim in this
Subsection and the Subsection to follow is to discuss the
renormalization of information in cases when the outcome space is
fractal or when the statistical system in question is
multifractal. Conclusions of such a reneormalization will be
applied is Subsection~\ref{can}.

%Let us begin with some fundamentals on fractals and multifractals.
Fractals are sets with a generally non--integer dimension
exhibiting property of self--similarity. The key characteristic of
fractals is fractal dimension which is defined as follows:
Consider a set $M$ embedded in a $d$--dimensional space. Let us
cover the set with a mesh of $d$--dimensional cubes of size $l^d$
and let $N_l(M)$ is a number of the cubes needed for the covering.
The fractal dimension of $M$ is then defined as~\cite{Man1,Fed1}
\begin{equation}
D = - \lim_{l \rightarrow 0} \frac{\ln N_l(M)}{\ln l}\, .
\label{fra1}
\end{equation}
\noi  In most cases of interest the fractal dimension (\ref{fra1})
coincides with the Hausdorff--Besicovich dimension used by
Mandelbrot~\cite{Man1}.

Multifractals, on the other hand, are related to the study of a
distribution of physical or other quantities on a generic support
(be it or not fractal) and thus provide a move from the geometry
of sets as such to geometric properties of distributions. Let a
support is covered by a probability of some phenomenon. If we pave
the support with a grid of spacing $l$ and denote the integrated
probability in the $i$th box as $p_i$, then the scaling exponent
$\alpha_i$ is defined~\cite{Man1,Fed1}
\begin{equation} p_i (l) \sim l^{\alpha_i}\, .
\end{equation}
%
%\noindent so the density
%
%\begin{equation}
%\varrho_i = \frac{p_i}{l^d} \propto l^{\alpha_i -d}\, .
%\end{equation}
%
The exponent $\alpha_i$ is called  singularity  or
Lipshitz--H\"{o}lder exponent.

Counting boxes $N(\alpha)$ where $p_i$ has $\alpha_i \in
(\alpha,\alpha + d\alpha)$, the singularity spectrum $f(\alpha)$
is defined as~\cite{Man1,Fed1}
\begin{equation}
N(\alpha) \sim l^{-f(\alpha)}\, .
\end{equation}
Thus a multifractal is the ensemble of intertwined (uni)fractals
each with its own fractal dimension $f(\alpha_i)$. For further
investigation it is convenient to define a ``partition
function"~\cite{Man1}
\begin{equation}
Z(q) = \sum_i p_i^q = \int d\alpha' \rho(\alpha') l^{-f(\alpha')}
l^{q\alpha'}\, .
\end{equation}
In the small $l$ limit the method of steepest descent yields the
scaling~\cite{Man1}
\begin{equation}
Z(q)\sim l^{\tau}\, , \label{tau}
\end{equation}
with
\begin{equation}
\tau(q) = \min_{\alpha} (q \alpha - f(\alpha)), \,\, f'(\alpha(q))
=q \, .
\end{equation}
This is precisely Legendre transform relation. So pairs
$f(\alpha), \alpha$ and $\tau(q), q$, are conjugates with the same
mathematical content.

Connection of R\'{e}nyi entropies with multifractals is frequently
introduced via generalized dimensions
\begin{eqnarray}
D_q = \lim_{l\rightarrow 0} \left( \frac{1}{(q-1)} \frac{\log
Z(q)}{\log l} \right) = -\lim_{l \rightarrow 0} {\cal{I}}_q
(l)/\log_2 l\, . \label{gendim1}
\end{eqnarray}
These have direct applications in chaotic
attractors~\cite{TCH1,MHJ1,KT1,HP1} and they also characterize,
for instance, intermittency of turbulence~\cite{TA1,GP1} or
diffusion--limited aggregates (DLA) like patterns~\cite{TCH2}. In
chaotic dynamical systems all $D_q$ are necessary to describe
uniquely e.g., strange attractors~\cite{HP1}. While the proof
in~\cite{HP1} is based on a rather complicated self--similarity
argumentation, by employing the information theory one can show
that the assumption of a self--similarity is not really
fundamental~\cite{Jiz1}. For instance, when the outcome space is
discrete then all $D_q$ with $q \in [1, \infty)$ are needed to
reconstruct the underlying distribution, and when
 the outcome space is $d$--dimensional subset of ${\IR}^d$
  then all $D_q$, $q \in (0,
\infty)$, are required to pinpoint uniquely the underlying PDF.
The latter examples are nothing  but the information theoretic
variants of Hausforff's moment problem of mathematical statistics.

%%%%%%%%%%%%%%%%%%%%%%%%%%%%%%%%%%%%%%%%%%%%%%%%%%%%%%%%%%%%%%%%%%%%%%%%%%%
\subsection{Fractals, multifractals and renormalization issue \label{fra2} }
%%%%%%%%%%%%%%%%%%%%%%%%%%%%%%%%%%%%%%%%%%%%%%%%%%%%%%%%%%%%%%%%%%%%%%%%%%%

In a close analogy with Section~\ref{Co1} it can be
shown~\cite{Jiz1} that for a fractal outcome space the following
asymptotic expansion of R\'{e}nyi's entropy holds
\begin{eqnarray}
{\mathcal{I}}_q(p_{kn}) \approx D \log_2n + h + o(1)\, ,
\label{exp1}
\end{eqnarray}
where $D$ corresponds to the Hausdorff dimension. The finite part
$h$ is, as before, chosen by the renormalization prescription -
additivity of information for independent experiments. Then
\begin{eqnarray}
{\cal{I}}_{q}({\cal{F}}) &\equiv& h  = \lim_{n \rightarrow \infty}
({\cal{I}}_{q}({\cal{P}}_n) - D \log_2 n) = \lim_{n \rightarrow
\infty} ({\cal{I}}_{q}({\cal{P}}_n) -{\cal{I}}_{q}({\cal{E}}_n) )
 \nonumber \\ &=& \frac{1}{(1-q)} \log_2 \left(
\int_M d \mu \, {\cal{F}}^{q}({\bf{x}}) \right)\, . \label{h1}
\end{eqnarray}
Measure $\mu$ in (\ref{h1}) is the Hausdorff measure
\begin{eqnarray}
\mu(d;l) = \sum_{\textrm{\scriptsize k-th box}} {l^d} \
\stackrel{l \rightarrow 0}{\longrightarrow} \ \left\{
\begin{array}{ll}
0 & \mbox{if $d < D$}\\
\infty & \mbox{if $d > D$}
\end{array} \right. \, .
\label{measure0}
\end{eqnarray}
Technical issues connected with integration on fractal supports
can be found, for instance, in~\cite{Ba1,Ed1}. Again, renormalized
entropy is defined as long as the integral on the RHS of
(\ref{h1}) exists.

We may proceed analogously with multifractals. The corresponding
asymptotic expansion now reads~\cite{Jiz1}
\begin{eqnarray}
{\mathcal{I}}_q(p_{nk}) \approx \frac{\tau(q)}{(1-q)} \log_2n + h
+ o(1)\, . \label{ren2}
\end{eqnarray}
This implies that
\begin{eqnarray}
h \equiv {\mathcal{I}}_{q}(\mu_{{\mathcal{P}}}) &=& \lim_{l
\rightarrow 0} \ \left({\mathcal{I}}_q({\mathcal{P}}_n) -
\frac{\tau(q)}{(q -1)} \log_2 n \right) = \lim_{l \rightarrow 0} \
\left({\mathcal{I}}_{q}({\mathcal{P}}_n) -
{\mathcal{I}}_{q}({\mathcal{E}}_n)  \right)
\nonumber \\
&=& \frac{1}{(1-q)} \ \log_2\left( \int_a
d\mu_{{\mathcal{P}}}^{(q)}(a)\right)\, . \label{h2}
\end{eqnarray}
Here the multifractal measure is defined as~\cite{Fed1}
\begin{eqnarray}
\mu_{{\mathcal{P}}}^{(q)}(d;l) = \sum_{\textrm{\scriptsize k-th
box}} \frac{p^{q}_{nk}}{l^d} \ \stackrel{l \rightarrow
0}{\longrightarrow} \ \left\{
\begin{array}{ll}
0 & \mbox{if $d < \tau(q)$}\\
\infty & \mbox{if $d > \tau(q)$}\, .
\end{array}
\right. \label{measure1}
\end{eqnarray}
It should be stressed that integration on multifractals is rather
delicate technical issue which is not yet well developed in the
literature~\cite{Ed1}.

%%%%%%%%%%%%%%%%%%%%%%%%%%%%%%%%%%%%%%%%%%%%%%%%%%%%%%%%%%%%%%%%%%
\subsection{Canonical formalism on multifractals \label{can}}
%%%%%%%%%%%%%%%%%%%%%%%%%%%%%%%%%%%%%%%%%%%%%%%%%%%%%%%%%%%%%%%%%%

We shall now present an important connection of R\'{e}nyi's
entropy with multifractal systems. The connection will be
constructed in a close analogy with canonical formalism of
statistical mechanics. As this approach is thoroughly discussed
in~\cite{Jiz1} we will, for shortness's sake, mention only the
salient points here.

Let us first consider a multifractal with a density distribution
$p(x)$. If we use, as previously, the covering grid of spacing $l$
then the coarse--grained Shannon's entropy of such a process will
be
\begin{equation}
{\mathcal{I}}({\mathcal{P}}_n(l)) = - \sum p_k(l) \log_2 p_k(l) \,
. \label{Shan1}
\end{equation}
Important observation of the multifractal theory is that when
$q=1$ then
\begin{eqnarray}
a(1) = \frac{d \tau(1)}{dq} = f(a(1)) = \lim_{l \rightarrow 0}
\frac{\sum_k p_k(l) \log_2 p_k(l)}{\log_2 l} = -\lim_{l
\rightarrow 0} \frac{{\mathcal{I}}({\mathcal{P}}_n(l))}{\log_2
l}\, , \label{Shan2}
\end{eqnarray}
describes the Hausdorff dimension of the set on which the
probability is concentrated - measure theoretic support. In fact,
the relative probability of the complement set approaches zero
when $l\rightarrow 0$. This statement is known as Billingsley
theorem~\cite{Bi1} or curdling~\cite{Man1}.

For the following considerations it is useful to introduce a
one--parametric family of normalized measures $\varrho(q)$
(zooming or escort distributions)
\begin{eqnarray}
\varrho_i(q,l) = \frac{[p_i(l)]^q}{\sum_j [p_j(l)]^q} \sim
l^{f(a_i)}\, . \label{zoom2}
\end{eqnarray}
Because
\begin{eqnarray} && df(a) = \left\{ \begin{array}{ll}
                          \leq da & \mbox{if $q\leq1$}\, ,\\
                          \geq da & \mbox{if $q\geq1$}\, ,
                          \end{array}\right.
%&\Rightarrow& \; \tau(q \leq 1) \leq 0\, , \;\;\; \mbox{and}
%\;\;\; \tau(q\geq 1) \geq 0 \, . \label{tau}
\label{zoom1}
\end{eqnarray}
\noi we obtain after integrating (\ref{zoom1}) from $a(q=1)$ to
$a(q)$ that
\begin{eqnarray}
&& f(a) = \left\{ \begin{array}{ll}
                          \leq a & \mbox{if $q\leq1$}\, ,\\
                          \geq a & \mbox{if $q\geq1$}\, .
                          \end{array}\right.
\end{eqnarray}
%\begin{eqnarray*}
%&&f(a_i)|_{q\leq1}\ \leq \ a_i - \tau(q\leq1) \ \leq \ a_i\, ,\nonumber \\
%&&f(a_i)|_{q\geq1}\ \geq \ a_i - \tau(q\geq1) \ \geq \ a_i\, ,
%\end{eqnarray*}
%
So for $q >1$ $\varrho(q)$ puts emphasis on the more singular
regions of ${\mathcal{P}}_n$, while for $q <1$ the accentuation is
on the less singular regions. Parameter $q$ thus provides a ``zoom
in" mechanism to probe various regions of a different singularity
exponent.

As the distribution (\ref{zoom2}) alters the scaling of original
${\mathcal{P}}_n$, also the measure theoretic support changes. The
fractal dimension of the new measure theoretic support
${\mathcal{M}}^{(q)}$ of $\varrho(q)$ is
\begin{eqnarray}
d_{h}({\mathcal{M}}^{(q)}) &=& \lim_{l\rightarrow 0}
\frac{1}{\log_2 l} \, \sum_k \varrho_k(q,l) \log_2
\varrho_k(q,l)\, . \label{measure2}
\end{eqnarray}
Note that the curdling (\ref{measure2}) mimics the situation
occurring in equilibrium statistical physics. There, in the
canonical formalism one works with (usually infinite) ensemble of
identical systems with all possible energy configurations. But
only the configurations with $E_i \approx \langle E(T )\rangle$
dominate at $n\rightarrow\infty$. Choice of temperature then
prescribes the contributing energy configurations.  In fact, we
may define the ``microcanonical" partition function as
\begin{equation}
Z_{mic} = \left(\sum_{a_k \in (a_i, a_i + d a_i)} \!\!\!\! 1
\right) = dN(a_i)\, . \label{partition1}
\end{equation}
Then the microcanonical (Boltzmann) entropy is
\begin{eqnarray}
{\mathcal{H}}({\mathcal{E}}(a_i)) = \log_2 dN(a_i) = \log_2
Z_{mic}\, , \label{boltzmann1}
\end{eqnarray}
and hence
\begin{eqnarray}
\frac{{\mathcal{H}}({\mathcal{E}}(a_i))}{\log_2 \varepsilon}\
\approx \ - \langle f(a) \rangle_{mic}\, . \label{f1}
\end{eqnarray}
Interpreting  $E_i = -a_i \log_2 \varepsilon$ as ``energy''we may
define the ``inverse temperature" $1/T = \beta /\ln2$ (note that
here $k_B = 1/\ln2 $) as
\begin{eqnarray}
1/T= \left.\frac{\partial {\mathcal{H}}}{\partial
E}\right|_{E=E_i} = -\frac{1}{\ln \varepsilon \ Z_{mic}}\
\frac{\partial Z_{mic}}{\partial a_i} = f'(a_i) =  q \, .
\label{temperature1}
\end{eqnarray}
On the other hand, with the ``canonical" partition function
\begin{equation}
Z_{can} = \sum_i p_{i}(\varepsilon)^{q} = \sum_i e^{-\beta E_i}\,
, \label{partition2}
\end{equation}
and $\beta=q \ln2$ and $E_i=-\log_2(p_i(\varepsilon)) $ the
corresponding means read
\begin{eqnarray}
&& a(q) \equiv \langle a \rangle_{can} = \sum_{i}
\frac{a_i}{Z_{can}} e^{-\beta E_i} \ \approx \  \frac{\sum_i
\varrho_i (q,\varepsilon)\log_2
p_i(\varepsilon)}{\log_2 \varepsilon} \, , \label{a1} \\
&& f(q) \equiv \langle f(a) \rangle_{can} = \sum_i
\frac{f(a_i)}{Z_{can}} e^{-\beta E_i}\ \approx  \ \frac{\sum_i
\varrho_i(q,\varepsilon) \log_2 \varrho_i(q,\varepsilon)}{\log_2
\varepsilon}\, . \label{f2}
\end{eqnarray}
Let us note particularly that the fractal dimension of the measure
theoretic support $d_h({\mathcal{M}}^{(q)})$ is simply $f(q)$. By
gathering the results together we have
\vspace{2mm}

\begin{center}
\begin{tabular}{|c|c|}  \hline\hline
{\bf{\,\, micro--canonical ensemble \,\,}}  & {\bf{
\,\, canonical ensemble \,\,}}  \\
{\bf{ - unifractals }}  & {\bf{
- multifractals }}\\
\hline ~&~\\$Z_{mic}$;
${\mathcal{H}} = S_{mic} = \log_2 Z_{mic}$ &
$\,\, Z_{can}; S_{can} = \log_2Z_{can} -q\langle a \rangle_{can} \log_2\varepsilon \,\,$  \\
~&~\\
 $\langle a \rangle_{mic} = a_i = \sum_k a_k /Z_{mic}$ & $\langle a \rangle_{can} =
 \sum_k a_k \ e^{-\beta E_k}/ Z_{can}$ \\
 ~&~\\
$\langle f(a) \rangle_{mic} = - S_{mic}/\log_2 \varepsilon $
& $\langle f(a) \rangle_{can} = - S_{can} / \log_2 \varepsilon $\\
~&~ \\
$q = \left. \partial S_{mic}/\partial E \right|_{E= E_i}$ &  $q =
\partial
S_{can}/\partial \langle E \rangle_{can}$ \\
~ & ~ \\
$\beta  = \ln 2/T = q$ & $\beta  = \ln 2/T = q$ \\
~ & ~ \\
$E_i = - \log_2 p_i = -a_i \log_2 \varepsilon$ & $\langle E
\rangle_{can} = -\langle a \rangle_{can} \log_2 \varepsilon$\\
~ & ~\\
$\langle f(a)\rangle_{mic} = q\langle a\rangle_{mic} - \tau$ &
$\langle
f(a)\rangle_{can} = q \langle a\rangle_{can} - \tau$\\
~ & ~ \\
 \hline\hline
\end{tabular}
\end{center}

\vspace{2mm}
%
%Comparison with the usual thermodynamics shows that the r\^{o}le
%of $n$ is played in multifractals by $-\log_2 \varepsilon$.
%
%The formal analogy with thermodynamics can be further illustrated
%by constructing an analogy to fluctuation--dissipation theorem.
%This might be done by realizing that
Looking at fluctuations of $a$ in the ``canonical" ensemble we can
establish an equivalence between unifractals and multifractls.
Recalling Eq.(\ref{tau}) and realizing that
\begin{eqnarray}
&&\partial^2 (\log_2 Z_{can})/\partial q^2 = \langle E^2
\rangle_{can}
- \langle E \rangle^2_{can} \approx (\log_2\varepsilon)^2 \, , \label{f-d0}\\
 &&\partial^2 (\tau \log_2 \varepsilon)/\partial q^2 = (\partial
 a/\partial q) \log_2 \varepsilon
\approx \log_2 \varepsilon \, , \label{f-d1}
\end{eqnarray}
we obtain for the relative standard deviation of ``energy"
\begin{eqnarray}
\frac{\sqrt{\langle E^2\rangle_{can} - \langle E\rangle^2_{can}}
}{\log_2 \varepsilon } = \sqrt{\langle a^2\rangle_{can} - \langle
a \rangle^2_{can}} \approx \frac{1}{\sqrt{-\log_2 \varepsilon }}
\rightarrow 0\, . \label{f-d2}
\end{eqnarray}
So for small $\varepsilon$ (i.e., exact multifractal) the
$a$--fluctuations become negligible
%\footnote{The Bilingsley
%theorem then can be viewed as a result of fluctuation--dissipation
%theorem.}
and almost all $a_i$ equal to $\langle a \rangle_{can}$.
If $q$ is a solution of the equation $a_i = \tau'(q)$ then in the
``thermodynamic" limit ($\varepsilon \rightarrow 0$) $a_i \approx
\langle a \rangle_{can}$ and the microcanonical and canonical
entropies coincide. Hence
\begin{eqnarray}
S_{mic}\approx - \sum_k \varrho_k(q,\varepsilon) \log_2
\varrho_k(q,\varepsilon) \nonumber \equiv
{\mathcal{H}}({\mathcal{P}}_n)\mbox{{$|$}}_{f(q)}\, . \label{clt1}
\end{eqnarray}
The subscript $f(q)$ emphasizes that the Shannon entropy
${\mathcal{H}}({\mathcal{P}}_n)$ is basically the entropy of an
unifractal specified by the fractal dimension $f(q)$. Legendre
transform then implies that
\begin{eqnarray}
{\mathcal{H}}({\mathcal{P}}_n)\mbox{{$|$}}_{f(q)} &\approx& -
qa(q) \log_2(\varepsilon) + (1-q){\mathcal{I}}_q ({\mathcal{P}})\,
. \label{legendre}
\end{eqnarray}
Employing the renormalization prescriptions (\ref{h1}) and
(\ref{h2}) we finally receive that
\begin{equation}
{\mathcal{I}}_q^r \ = \ {\mathcal{H}}^r\mbox{\large{$|$}}_{f(q)}\,
. \label{result}
\end{equation}
So by changing the $q$ parameter R\'{e}nyie's entropy ``skims
over" all renormalized unifractal Shannon's entropies. R\'{e}nyi's
entropy thus provides a unified information measure which keeps
track of all respective unifractal Shannon entropies.

The passage from multifractals to single--dimensional statistical
systems is done by assuming that the $a$--interval gets
infinitesimally narrow and that PDF is smooth. In such a case both
$a$ and $f(a)$ collapse to $a = f(a) \equiv D$ and $q = f'(a) =1$.
For instance, for a statistical system with a smooth measure and
the support space ${\IR}^d$ Eq.(\ref{result}) constitutes a
trivial identity. We believe that this is the primary reason why
Shannon's entropy plays such a predominant role in physics of
single--dimensional sets. Discussion of (\ref{result}) can be
found in~\cite{Jiz1}.

\subsection{R\'{e}nyi's entropy and Fisher's information \label{fish}}
%%%%%%%%%%%%%%%%%%%%%%%%%%%%%%%%%%%%%%%%%%%%%%%%%%%%%%%%%%%%%%%%%%%%%%%%%
%
Let us present here an interesting connection which exists between
Riemaniann geometry on statistical parameter spaces and R\'{e}nyi
entropies.

Consider a family of PDF's characterized by a vector parameter
${\mathbf{\theta}}$
\begin{eqnarray}
{\mathcal{F}}_\theta = \left\{p(x, \theta); x \in M; \theta \in
{\mathcal{M}}, \mbox{a manifold in} \, {\IR}^n \right\}\, .
\label{param1}
\end{eqnarray}
We further assume that $p(x, \theta) \in$  ${\mathcal{C}}^2$. The
Gibbs PDF's (with $\theta_i$ being the inverse temperature
$\beta$, the external field $H$, etc.) represent example of
(\ref{param1}).

To construct a metric on ${\mathcal{M}}$ which reflects the
statistical properties of the family (\ref{param1}) Rao a
co-workers~\cite{Rao1} proposed to adopt various measures of
dissimilarity between two probability densities, and then use them
to derive the metric. Important class of dissimilarity measures
are measures based on information theory. Typically it is utilized
the gain of information when a density $p(x,\phi)$ is replaced
with a density $p(x,\theta)$. In the case of R\'{e}nyi's entropy
this is~\cite{Re1}
\begin{equation}
{\mathcal{I}}_q(\theta|| \phi) = \frac{1}{q-1} \log_2\int_M dx \
\frac{p(x, \theta)^q}{p(x, \phi)^{q-1}}\, .
\end{equation}
The information metric on ${\mathcal{M}}$ is then defined via the
leading order of dissimilarity between $p(x,\theta)$ and
$p(x,(\theta + d\theta))$, namely
\begin{equation}
{\mathcal{I}}_q(\theta ||\theta + d\theta) = \frac{1}{2!}
\sum_{i,j} {{\textsl{g}}}_{ij}(\theta) \ d\theta_i d\theta_j +
\ldots \, . \label{gij}
\end{equation}
Note that because ${\mathcal{I}}_q(\theta ||\phi)$ is minimal at
$\theta = \phi$, linear term in (\ref{gij}) vanishes. So we have
\begin{eqnarray}
{{\textsl{g}}}_{ij}(\theta) &=& \left[\frac{\partial^2 }{\partial
\phi_i
\partial \phi_j} {\mathcal{I}}_q(\theta|| \phi) \right]_{\theta =
\phi} = \frac{q}{2 \ln 2}\left( \int_M dx \ p(x,\theta)
\frac{\partial \ln p(x, \theta)}{\partial \theta_i}\frac{\partial
\ln p(x, \theta)}{\partial \theta_j} \right)\nonumber \\ &=&
\frac{q}{2 \ln 2} \ F_{ij}(p(x,\theta)\, .
\end{eqnarray}
Here $F_{ij}$ is the Fisher information matrix (or Fisher--Rao
metric)~\cite{Frie1,Rao}. Fisher matrix is the only Riemaniann
metric which is invariant under transformation of variables as
well as reparametrization~\cite{Rao}. In addition, the diagonal
elements of Fisher's information matrix represent the amount of
information on $\theta_i$ in an element of a sample chosen from a
population of density functions $p(x,\theta)$. Due to its relation
with Cram\'{e}r--Rao inequality Fisher information matrix plays a
crucial r\^{o}le in parametric estimation~\cite{Frie1}. Let us
stress that the latter is used in quantum mechanics to formulate
information uncertainty relations~\cite{Frie1,HM1}.

%%%%%%%%%%%%%%%%%%%%%%%%%%%%%%%%%%%%%%%%%%%%%%%%%%%%%%%%%%%%%%%%%%%%%%%%%
\section{Tsallis' entropy \label{Tsa}}
%%%%%%%%%%%%%%%%%%%%%%%%%%%%%%%%%%%%%%%%%%%%%%%%%%%%%%%%%%%%%%%%%%%%%%%%%

Tsallis' entropy (or non--extensive entropy, or $q$--order entropy
of Havrda and Charv\'{a}t~\cite{Rao,Havrda}) has been recently
much studied in connection with  long--range correlated systems
and with non--equilibrium phenomena. Although firstly introduced
by Havrda and Charv\'{a}t in the cybernetics theory
context~\cite{Havrda} it was Tsallis and co--workers~\cite{Ts1}
who exploited its non--extensive features and placed it in a
physical setting. Applications of Tsallis' entropy are ranging
from 3--dimensional fully developed hydrodynamic turbulence,
2--dimensional turbulence in pure electron plasma, Hamiltonian
systems with long--range interactions to granular systems and
systems with strange non--chaotic attractors. The explicit form of
Tsallis' entropy reads
\begin{equation}
{\cal{S}}_{q} = \frac{1}{(1-q)}\left[\sum_{k=1}^{n}(p_k)^{q}
-1\right] \, , \;\;\;\; q > 0 \, . \label{tsallis1}
\end{equation}
This form indicates that ${\mathcal{S}}_q$ is a positive and
concave function in ${\mathcal{P}}$. In the limiting case
$q\rightarrow 1$ one has $\lim_{q \rightarrow 1}{\cal{S}}_{q} =
\lim_{q \rightarrow 1}{\cal{I}}_{q} ={\mathcal{H}}$. In addition,
(\ref{tsallis1}) obeys a peculiar non--extensivity rule
\begin{equation}
{\cal{S}}_{q}(A \cap B)= {\cal{S}}_{q}(A) + {\cal{S}}_{q}(B|A) +
(1-q){\cal{S}}_{q}(A){\cal{S}}_{q}(B|A)\, . \label{tsallis2}
\end{equation}
It might be proved that the following axioms uniquely specify
Tsallis' entropy~\cite{Ab2}:

\vspace{3mm}
\begin{enumerate}
\item For a given integer $n$ and given ${\cal{P}} = \{ p_1, p_2,
\ldots , p_n\}$ ($p_k \geq 0, \sum_k^n p_k =1$),
${\cal{S}}({\cal{P}})$ is a continuous with respect to all its
arguments.

\item For a given integer $n$, ${\cal{S}}(p_1, p_2, \ldots , p_n)$
takes its largest value for $p_k = 1/n$ ($k=1,2, \ldots, n$).
%${\cal{I}}_{\alpha}({\cal{P}})$ takes its largest value for
%$p_k = 1/n, (k = 1,2, \ldots, n)$, i.e., the gained information is
%largest when we known least about the original system.

\item For a given $q\in {\IR}$; ${\mathcal{S}}(A\cap B) =
{\mathcal{S}}(A) +
{\mathcal{S}}(B|A) + (1-q){\mathcal{S}}(A){\mathcal{S}}(B|A)$ with\\
\\
 $\mbox{\hspace{2.5cm}}{\mathcal{S}}(B|A) =
\sum_k \varrho_k(q) \ {\mathcal{S}}(B|A=A_k)$,\\
\\
and $\varrho_k(q) = p_k^q/\sum_k p_k^q$\, (distribution
${\mathcal{P}}$ corresponds to the experiment $A$).

%For independent events, i.e., ${\cal{R}} = {\cal{P}}\times
%{\cal{Q}}=
%\{p_iq_k \}$: \\
%${\cal{I}}_{\alpha}({\cal{P}}\times {\cal{Q}}) =
%{\cal{I}}_{\alpha}({\cal{P}}) + {\cal{I}}_{\alpha}({\cal{Q}})$.
%
%\item In general case there is a continuous invertible function
%$f(x)$ such that: ${\cal{I}}_{\alpha}\partial
%f({\cal{I}}_{\alpha})/\partial
%{\cal{I}}_{\alpha} = g({\cal{I}}_{\alpha}) f({\cal{I}}_{\alpha})$.\\
%\\
%Here ${\cal{I}}_{\alpha}({\cal{P}}\cup {\cal{Q}}) =
%{\cal{I}}_{\alpha}({\cal{I}}_{\alpha}({\cal{P}}),
%{\cal{I}}_{\alpha}(\cal{Q}))$. The scaling function $g(x)$ is
%common to both ${\cal{I}}_{\alpha}({\cal{P}}\cup {\cal{Q}}),
%{\cal{I}}_{\alpha}({\cal{P}})$ and ${\cal{I}}_{\alpha}({\cal{Q}})$
%.

\item ${\cal{S}}(p_1,p_2, \ldots , p_n, 0 ) = {\cal{S}}(p_1,p_2,
\ldots , p_n)$.
\end{enumerate}

Note that these axioms bear remarkable similarity to the axioms of
R\'{e}nyi's entropy. Only the axiom of additivity is altered. We
keep here the linear mean but generalize the additivity law. In
fact, the additivity law in axiom~3 is the Jackson sum (or
$q$--additivity) of $q$--calculus. The Jackson basic number
$[X]_q$ of quantity $X$ is defined as $[X]_q = (q^X -1)/(q-1)$.
This implies that for two quantities $X$ and $Y$ the Jackson basic
number $[X+Y]_q = [X]_q + [Y]_q + (q-1)[X]_q[Y]_q$. The connection
with axiom~3 is established when $q \rightarrow (2-q)$.

The former axiomatics might be viewed as the $q$--deformed
extension of Shannon's information theory. Obviously, in the
$q\rightarrow 1$ limit the Jackson sum reduces to ordinary
${\mathcal{S}}(A\cap B) = {\mathcal{S}}(A) + {\mathcal{S}}(B|A)$
and above axioms boil down to Shannon--Khinchin axioms of
classical information theory~\cite{Kh1}.

Emergence of $q$--deformed structure allows to formalize many
calculations. For instance, using the $q$--logarithm~\cite{Ab3},
i.e., $\ln_q x = (x^{1-q} -1)/(1-q)$,  Tsallis' entropy
immediately equals to the $q$--deformed Shannon's entropy (again
after $q \rightarrow (2-q)$ ), i.e.
\begin{equation}
{\mathcal{S}}_q({\mathcal{P}}) = - \sum_{k=1}^n p_k \ln_q p_k\, .
\end{equation}
The interested reader may find some further applications of
$q$--calculus in non--extensive statistics, e.g., in~\cite{Ab3}

Let us finally add a couple of comments. Firstly, it is possible
to show~\cite{Jiz1} that R\'{e}nyi's entropy prescribes in a
natural way the renormalization for Tsallis' entropy in cases when
the PDF is absolutely continuous. This might be achieved by
analytically continuing the result for renormalized R\'{e}nyi
entropy from the complex neighborhood of $q=1$ to the entire right
half of the complex plane~\cite{Jiz1}. Thus, if ${\mathcal{F}}$ is
the corresponding PDF then
\begin{eqnarray*}
{\mathcal{S}}_q ({\mathcal{F}}) \equiv \lim_{n \rightarrow \infty
} \left(\frac{{\mathcal{S}}_q ({\mathcal{P}}_n)}{n^{D(1-q)}} -
\frac{{\mathcal{S}}_q ({\mathcal{E}}_n)}{n^{D(1-q)}} \right) =
\frac{1}{(1-q)} \, \int_M d\mu \,
{\mathcal{F}}({\mathbf{x}})\left(
{\mathcal{F}}^{q-1}({\mathbf{x}}) - 1\right)\, . \label{tsallis3}
\end{eqnarray*}
Extension to multifratals is more delicate as a possible
non--analytic behavior of $f(a)$  and $\tau(q)$ invalidates the
former argument of analytic continuation. These ``phase
transitions'' are certainly an interesting topic for further
investigation.

Secondly, note that Tsallis and R{\'e}nyi's entropies are
monotonic functions of each other and thus both are maximized by
the same ${\mathcal{P}}$. This particularly means that whenever
one uses MaxEnt approach (e.g., in thermodynamics, image
processing or pattern recognition) both entropies yield the same
results.

\vspace{-1mm}
%%%%%%%%%%%%%%%%%%%%%%%%%%%%%%%%%%%%%%%%%%%%%%%%%%%%%%%%%%%%%%%%%%%%%%%%%
\section{Conclusions \label{Con}}
%%%%%%%%%%%%%%%%%%%%%%%%%%%%%%%%%%%%%%%%%%%%%%%%%%%%%%%%%%%%%%%%%%%%%%%%%

In this lecture we have reviewed some information--theoretic
aspects of generalized statics of R\'{e}nyi and Tsallis.

Major part of the lecture - Section~2 - was dedicated to
R\'{e}nyi's entropy. We have discussed the information--theoretic
foundations of R\'{e}nyi's information measure and its
applicability to systems with continuous PDF's. The latter include
systems with both smooth (usually part of $ {\IR}^d$) and fractal
sample spaces. Particular attention was also paid to currently
much studied multifractal systems. We have shown how the R\'{e}nyi
parameter $q$ is related to multifractal singularity spectrum and
how R\'{e}nyi's entropy provides a unified framework for all
unifractal Shannon entropies.

In cases when the physical system is described by a parametric
family of probability distributions one can construct a Riemannian
metric over this probability space - information metric. Such a
metric is then a natural measure of diversity between two
populations. We have shown that when one employs R\'{e}nyi's
information measure then the information metric turns out to be
Fisher--Rao metric (or Fisher's information matrix).

In Section~3 we have dealt with Tsallis entropy. Because detailed
discussions of various aspects of Tsallis statistics are presented
in other lectures of this series we have confined ourselves to
only those characteristics of Tsallis' entropy which make it
interesting from information--theory point of view. We have shown
how the $q$--additive extension of original Shannon--Khinchin
postulates of information theory gives rise to $q$--deformed
Shannon's measure of information - Tsallis entropy.
%We have
%provided also some comments on connection of Tsallis entropy with
%information metric.

\vspace{-1mm}
%%%%%%%%%%%%%%%%%%%%%%%%%%%%%%%%%%%%%%%%%%%%%%%%%%%%%%%%%%%%%%%%%%%%%%%%%%%
\subsection*{Acknowledgements}
%%%%%%%%%%%%%%%%%%%%%%%%%%%%%%%%%%%%%%%%%%%%%%%%%%%%%%%%%%%%%%%%%%%%%%%%%%%

We would like to thank to organizers of the Piombino workshop
DICE2002 on ``Decoherence, Information, Complexity and Entropy"
for their kind invitation. It is a pleasure to thank to Prof
C.~Tsallis and Dr A.~Kobrin for reading a preliminary draft of
this paper, and for their comments. The work was supported by the
ESF network COSLAB and the JSPS fellowship.

%INDEX%%%%%%%%%%%%%%%%%%%%%%%%%%%%%%%%%%%%%%%%%%%%%%%%%%%%%%%%%%%%%%%
% Please check with the editor of your book whether he plans to
% include a "mutual" subject index - if so, please code your entries
% in the standard syntax. For your own purposes you may print your
% "personal" index by using the following commands:
%
%\clearpage
%\addcontentsline{toc}{section}{Index}
%\flushbottom
%\printindex
%%%%%%%%%%%%%%%%%%%%%%%%%%%%%%%%%%%%%%%%%%%%%%%%%%%%%%%%%%%%%%%%%%%%%

\end{document}